\documentclass[fintech,article,submit,moreauthors]{Definitions/mdpi}

\usepackage{graphicx}
\usepackage{tablefootnote}

\usepackage{amsmath, amssymb, amsthm}
\usepackage{tikz-cd}

\tikzset{
  symbol/.style={
    draw=none,
    every to/.append style={
      edge node={node [sloped, allow upside down, auto=false]{$#1$}}}
  }
}

\newtheorem{observation}{Observation}

\usepackage{tabularx,makecell,xcolor}
\newcolumntype{Y}{>{\raggedright\arraybackslash}X}
\newcolumntype{P}[1]{>{\raggedright\arraybackslash}p{#1}} 
\newcommand{\Main}{\textcolor{green!60!black}{\Large$\bullet$}}
\newcommand{\Tang}{\textcolor{yellow!80!black}{\Large$\bullet$}}
\newcommand{\None}{\textcolor{gray}{\Large$\circ$}}

\firstpage{1} 
\pubvolume{1}
\issuenum{1}
\articlenumber{0}
\pubyear{2026}
\copyrightyear{2026}
\externaleditor{Firstname Lastname} 
\datereceived{23 July 2026} 
\daterevised{21 September 2026} 
\dateaccepted{15 October 2026} 
\datepublished{ } 

\Title{Perspectives on Objectives and Design Principles in Offline Payments with Central Bank Digital Currency}

\Author{David-Alexandre Guiraud $^{1}$\orcidA{}, Andrea Tundis $^{1,2}\orcidB{}$ and Marc Winstel $^{1,}$*\orcidC{}} 

\AuthorNames{David-Alexandre Guiraud, Andrea Tundis and Marc Winstel}

\address{%
$^{1}$ \quad Directorate-General Digital Euro, Deutsche Bundesbank, {60325} Frankfurt am Main
, Germany; david-alexandre.guiraud@bundesbank.de (D.-A.G.); andrea.tundis@bundesbank.de {(A.T.)} 
\\
$^{2}$ \quad  Department of Computer Science, Technische Universität Darmstadt, {64289} Darmstadt, Germany}

\corres{Correspondence: marc.winstel@bundesbank.de}

\abstract{In this work, fundamental design principles for a \textit{central bank digital currency} 
(CBDC) with offline functionality and corresponding countermeasures are discussed.
    We identify three major objectives relevant for core security and privacy properties of such CBDC proposals:
    (i) \textit{Access Control Security}---
protection of a user's funds against unauthorized access by other users; (ii) \textit{Security against Depositor Misbehavior}---preservation of the integrity of an  environment (potentially the wallet) against misbehavior of its owner (for example, double-spending); (iii) \textit{Privacy by Design}---ensuring privacy is embedded into the system architecture. 
    Our central conclusion is based on the alignment of the objectives to concrete design elements as countermeasures; certain objectives and countermeasures have no or minimal interference with each other. 
    A major consequence of our work 
    is that the integrity of a user's wallet and, accordingly, the mitigation of double-spending race attacks should be addressed through the adaptation and integration of \textit{secure hardware} or functionally equivalent trusted execution mechanisms within a CBDC system. 
    The developed framework is applied for the analysis of five academic proposals of \mbox{payment systems}.}

\keyword{\textls[-15]{central bank digital currency; offline payments; digital banking; double-spending;} CBDC system design} 

\begin{document}
\Crefname{observation}{observation}{observations} 
\Crefname{observation}{Observation}{Observations}
\nolinenumbers

\setcounter{footnote}{0}

\section{Introduction}\label{sec:introduction}

In recent years, central banks have explored and worked on a new form of digital money, known as \textit{central bank digital currencies} (CBDCs), issued by central banks.
Thus, one typically distinguishes between wholesale CBDCs, specifically intended for transactional purposes among central banks and financial institutions, where access is typically limited to these institutions ({note} that wholesale CBDCs, in~principle, already exist in the form of bank accounts that commercial banks and other institutions hold at their respective central bank; however, in~many contexts, one refers with the term wholesale CBDC to novel developments for wholesale purposes like distributed ledger technologies and/or tokenized central bank money.)
, and~retail CBDCs, which are designed for e-commerce payments, for~retail (point of sale) payments in physical shops, or~for payments between individuals (peer to peer).
The latter are sometimes referred to as a ``digital equivalent of cash'' (cf. the definition in~\cite{repec:bis:bisbps:159}: ``A retail CBDC is a digital version of physical cash that households and firms can use for everyday transactions. A~retail CBDC differs from existing forms of cashless payment instruments, such as credit transfers, direct debits, card payments and e-money, as~it is a direct liability of the central bank rather than a liability of a private financial 
institution.''). 
\textit{Offline payment functionalities} (OPFs) are an important but complex feature of retail CBDCs, which can, for instance, extend their accessibility and more closely resemble the properties of~cash. 

This article discusses the challenges that arise in the design of a retail CBDC with OPF~\cite{IMFOfflineCBDCBlueprint,quantnetworkoffline}, particularly regarding security and privacy. 
These challenges are investigated in a generic setup, i.e.,~with only minimal specification of the particular implementation.
The definition of this setup is part of \Cref{sec:problemdefinition}.

Practically all proposed CBDC schemes with OPF try to achieve security and privacy goals by making use of cryptographic methods, \textit{secure hardware} and/or \textit{Zero-knowledge Proof} (ZKP) techniques; see~\cite{chu2022offline} for a review and~\cite{Atangana23,kempen2024offlinedigitaleurominimum,beer2024payoffregulatedcentralbank} for more recent literature.
In this work, we first of all identify particularly relevant security and privacy objectives for CBDCs with an~OPF.

\begin{figure}[H]
    
    \includegraphics[width=0.95\linewidth]{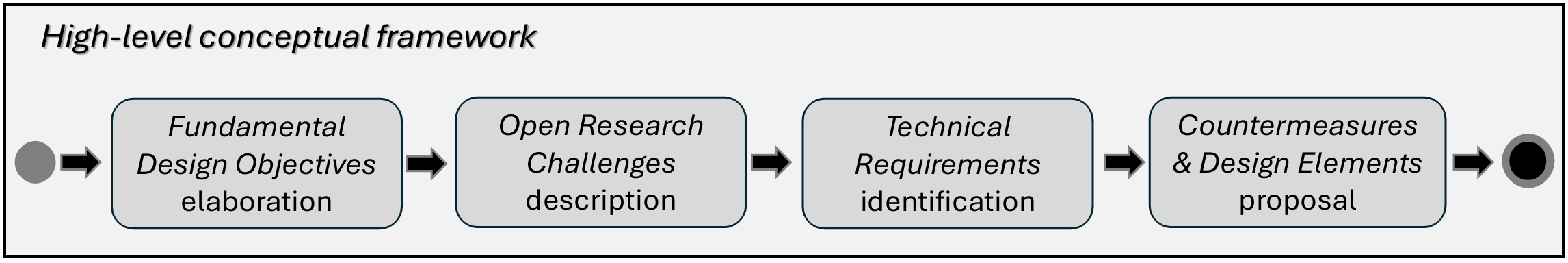}
    \caption{The framework provides a structured approach organized in 4 main conceptual components for assessing objectives and design principles in offline payments with central bank digital currency (CBDC). First, the fundamental design objectives that a general offline CBDC payment system (as defined in \cref{sec:problemdefinition}) should satisfy are addressed; then, the open research challenges that may hinder their achievement are identified. Based on these challenges, the~corresponding technical requirements are derived and, finally, it deals with the countermeasures and design elements to address the identified requirements in \cref{sec:AlignObjectivesAndDesignElements}. Such components are interconnected, providing a  path from high-level objectives to concrete technical and design~considerations.}
    \label{fig:conceptual-framework}
\end{figure}

Based on these objectives, we investigate what contribution each 
of the aforementioned design elements can make to offline CBDC design.  
Rather than studying security and privacy mechanisms in isolation, this article analyzes their specific roles within the threat structure of offline CBDCs. As~a central result, we derive an association between objectives, design elements, countermeasures, and~their limitations.  
Such a concrete alignment can be implemented when designing a CBDC protocol, i.e.,~in a way that there is only little or no interference at all between different design elements regarding the identified objectives. An~overview of the high-level framework is depicted in Figure~\ref{fig:conceptual-framework}.
 
One of the main upshots of our discussion is that, under~the assumptions of transferable offline payments defined in \Cref{def:transferabilityoffline}, the~possibility of double-spending (later in this work, the~problem of double-spending is properly defined) during an offline period is a structural risk.
In particular, real-time validation against the TA is unavailable at the moment of transfer, so the CBDC system cannot
ensure ledger consistency on a global~level.

This constitutes the essence of the difficulty in designing a CBDC with transferable OPF. 
Given the threat model considered here, one can at least attempt to make double-spending difficult and detectable by shielding the
wallet's execution environment and memory against manipulation by its legitimate user.
One prominent proposal for doing this is via \textit{secure hardware} (e.g., Secure Elements) or functionally equivalent trusted execution mechanisms. Hence, a~more thorough understanding of the possibilities and limitations of trusted hardware in its own sense is necessary. 
However, we want to point out that these mechanisms provide local protection mechanisms, but~do not by themselves lead to common ledger~consistency.

This also leads to the important observation that introducing an OPF into a currency area's CBDC program inherently introduces the manufacturer(s) and supplier(s) of trusted hardware as additional trust anchors (TAs) in addition to the trusted party running a ledger or other database (usually the central bank or a group permissioned by the central bank). 
This makes it all the more critical and important that privacy is embedded into the system by design, so that users do not have to rely solely on institutional or technical TA but can be assured that their personal data is protected at the architectural~level.

Several publications provide frameworks for structuring the design space of retail CBDCs. Ref.~\cite{auer2020technology} organizes CBDC design around a hierarchy of consumer needs, technical design choices, and~central-bank architecture. More detailed taxonomies of CBDC design decisions and possible system architectures are developed in~\cite{stamm2023taxonomy, tsareva2024retail}. 
With respect to offline payments specifically, the~BIS Project Polaris report~\cite{Polaris} provides comprehensive guidance on relevant design choices, risks, and~implementation considerations, and~\cite{chu2022offline} reviews offline-CBDC approaches from the perspective of security requirements. 
More recently, ref.~\cite{nekriach2025assessment} proposes a simulation-based assessment framework for evaluating the resilience of offline digital-currency systems under different operating conditions and levels of malicious activity. 
These contributions provide valuable taxonomies, requirements, and~evaluation methodologies. 
However, a~derivation of a set of objectives for the security and privacy of offline CBDC systems based on a generic model of an offline-capable CBDC system, as well as a systematic relation of these objectives to concrete architectural countermeasures, is currently missing in the~literature.

Our work addresses this gap by identifying such objectives and using them as a technology-independent basis for reasoning about offline-CBDC design.
Methodologically, this paper aims at a conceptual and design-oriented analysis rather than an empirical or experimental study. Starting from a generic model of an offline-capable CBDC system, we derive a threat classification, identify central security and privacy objectives, and~develop an assignment between these objectives and corresponding design elements and countermeasures. The~resulting framework is then applied to selected academic CBDC proposals in order to illustrate its use for comparative~analysis.

In \Cref{sec:problemdefinition}, CBDC systems with OPFs and their properties, as well as common security threats, are defined with minimal specifications regarding the implementation.
Then, \Cref{sec:Privacy} discusses privacy concepts in a general manner.
Thus, we identify \textit{Privacy by Design} as a proper leitmotif for CBDCs in general, and~discuss methods to implement \textit{Privacy by Design} for CBDCs. 
In \Cref{sec:AlignObjectivesAndDesignElements}, three main objectives for privacy and security in CBDCs with OPF are formulated and an alignment of objectives with concrete design elements is identified.
Implications of this analysis for the assessment and construction of offline CBDCs are discussed.
Before concluding in \Cref{sec:conclusion}, we apply the framework illustratively to examples from contemporary literature in \Cref{sec:proposal_examination}.

\section{Definition of Concepts and Problem~Description} \label{sec:problemdefinition}

As stated earlier, we focus on CBDCs for retail purposes, i.e.,~CBDCs intended to cater to the needs of individuals or households for everyday transactions.
Such retail CBDCs are often envisioned as a digital central bank liability with cash-like properties.
Besides the promise of maintaining a stable value as central bank money, the usually high standards with respect to security and privacy are expected from any CBDC~implementation.

Research around CBDCs is a rather novel and evolving field, such that the terminology of the field is not yet standardized.
For the sake of formulating design objectives and principles in this work, we therefore provide definitions for key concepts relevant to  a retail CBDC with OPF.
Therefore, we summarize the core elements of a retail CBDC as follows:
\begin{Definition}\label{def:cbdc}
    A CBDC is a financial instrument based on a digital payment infrastructure where values are represented as direct liabilities of the central bank(s) {(the central bank itself is the ultimate debtor, that means, it is directly responsible for honoring the value represented)}, comprising the following~elements:
    \begin{itemize}
    \item Users, each capable of maintaining digital wallets (one or multiple) with the purpose of safely storing and transferring funds;
     
     \item A Trust Anchor (TA), a~designated institution which acts as a universally trusted reference for the users (for example, a~central bank or a group of permissioned entities could fulfill this role);

    \item A sufficiently  detailed set of specifications, e.g.,~rulebooks, protocol specifications, pseudocode implementations etc., defining the rules and mechanisms by which users can securely store and transfer their funds to other users within the system;

    \item Optionally, 
        intermediary entities facilitating access to the system and/or the processing of transactions. 
    \end{itemize}
\end{Definition}

\begin{Definition}
    In this context, we understand by a \textbf{fund} 
    a liability against
    the central bank (usually denominated in units of 
    the country's base currency) that can be held only by a single, unique user.
    In this sense,
    a \textbf{transfer of funds} is a consensual displacement of such funds between two participants, from~a sender to a receiver, adhering to the specifications of the CBDC system.
    A transfer is considered \textbf{final} as soon as it cannot be reversed or invalidated by the sender (even if the sender is not adhering to the specifications of the CBDC during his attempt to reverse the transfer).
\end{Definition}

Observe that in this definition, finality is a global notion, i.e.,~we can assess finality of a transfer only if we know the full state of the CBDC system.
Note that we use finality of a transfer below in a \textit{subjective sense} as well, i.e.,~a user considers the receipt of funds as final if he is assured to a sufficiently high degree by the information available to him that the credited amount cannot be revoked later on. {(Note that only the subjective meaning of finality is typically achieved in everyday life. Even when transactions are processed using cash, objective finality can only be achieved when the receiver is able to check for counterfeit money concurrently with the transfer of funds.)}
In the sense of \Cref{def:cbdc}, intermediary entities could be commercial banks or other commercial payment service providers, whose particular role depends on the specific design of a CBDC and is not further discussed in this~manuscript.

\par{\textbf{Abstract scheme for transactions.}}
Each CBDC system needs to define a certain protocol {(we remark that the protocol usually forms only a small (technical) portion of a more
general rulebook; in~such a rulebook, a~comprehensive set of rules are laid out in order to specify the overall CBDC functioning as well as interactions with the broader ecosystem in which the CBDC is meant to operate)} for the initiation and settlement of transactions.
We assume that the CBDC systems analyzed in the remainder of this article implement protocols that follow the abstract scheme outlined here:
A transaction is initiated by the sender---possibly with participation of the receiver---who constructs and transmits a suitable dataset in a formally defined way and transmits it to designated CBDC entities (usually, the~receiver and/or the TA) to notify them of the intended transfer and to initiate the necessary settlement or accounting steps.
Such a data set contains at least (possibly in an indirect way and/or in potentially anonymized or pseudonymized form):
\begin{itemize}
    \item A specification of the funds to be transferred; 
    \item A specification of the wallets or users involved in the transfer (typically sender specification and receiver specification). 
\end{itemize}

While 
 the technical realization and structure of such a transaction can drastically vary depending on the particular implementation, it is essential that this information must be extractable by verifying entities from the data set and communication metadata (between sender and receiver) in order to check validity and perform further processing.
In many protocol proposals, this information is given as a transaction statement, which specifies the above data explicitly.
In the following, we use the generic term transaction statement for the information contained in this~transaction.

Many proposed CBDC systems require that the receiver can communicate with the TA (possibly relayed by the sender) in order to check settlement before handing out the exchange value of the transaction. 
Such a system is regarded as an \textit{online CBDC}.

In many proposed or prototyped CBDC systems~\cite{tang2025central}, the~TA maintains a (centralized or suitably distributed) ledger, 
a database which acts as a universally accepted reference.
In this sense, in contrast to permissionless cryptocurrencies, CBDC systems typically do not aim to achieve decentralization through consensus mechanisms.  
We remark that this distinction concerns primarily institutional governance rather than the technical or infrastructural architecture. The~TA represents an institutional trust role without, by~itself, implying a particular infrastructural design choice, such as a centralized database. Accordingly, centralized governance by a central bank or other authorized entities could be implemented through distributed databases or permissioned DLT models. Our work therefore does not rule out the existence of such alternative solutions, such as, e.g.,~discussed in~\cite{bindseil2026tokenizing}. Rather, we adopt centralized governance as a working assumption for maintaining monetary integrity instead of assuming alternative permissionless consensus~mechanisms.

Then, it is often stipulated that any transaction must eventually be validated and 
incorporated into this ledger by the TA. 
A user---usually we consider the receiving user of the transaction in question---can be asserted to a high degree so that the transaction will not be revoked and can be used for further spending if the transaction is settled in the central ledger, i.e.,~the user considers this transaction as~final.

We already mentioned that technical details of the CBDC system, such as the data representation of funds, will not be essential for the discussion below. 
On a more abstract level, however, it is important for the following discussion to distinguish between CBDC systems designed as Bearer instruments and account-based CBDCs.
The latter reminds of traditional bank accounts with the TA maintaining accounts in the ledger, i.e.,~the actual funds lie in the ledger of the TA.
A bearer instrument CBDC implementation behaves more like physical cash in the sense that users hold the digital representation of the fund within their respective wallet without a third party having any control over these~funds.

\begin{Definition}\label{def:offlinePayment}
    An \textbf{offline payment functionality (OPF)} allows to transfer funds within the CBDC while the two parties (payer and payee) of the transaction can only communicate with each other (e.g., via NFC, Bluetooth or QR-Codes), but~not with any other entity.
\end{Definition}

This usually amounts to a sender forming the transaction statement
and providing a suitable authorization of himself
and/or the funds involved, whereas the receiver is limited to a local validity check
of some sort. 
As a consequence, payer and payee must be able to arrange a transaction to the point where both consider it final without relying on the trust provided by the~TA.

We remark that offline payments differ fundamentally from online transactions with regard to system integrity:
In an offline setting, the~mechanism of updating a common ledger is not available at the time of payment, since the payer and payee may not be connected to the central infrastructure or network.
Consequently, the~OPF requires a different integrity model based, e.g.,~on hardware security assumptions.
Thus, our analysis does not consider OPF as an add-on to a generic blockchain mechanism in which decentralization or distributed consensus are main principles. 
In a CBDC system, centralized governance and an authoritative reference maintained by the TA are fundamental assumptions, as~the currency ultimately remains a liability of the central bank. 
The relevant question is therefore how offline payments can be enabled within this centrally governed setting, and~how resulting risks can be controlled without relying on real-time~consensus.  

An important distinction is that between the integrity of the local execution environment and the consistency of the common ledger. Secure hardware mainly protects the wallet state and payment protocol logic, thereby contributing to execution integrity. Ledger consistency, on~the other hand, refers to a consistent global view of the CBDC state and can be achieved by the TA or through an appropriate consensus mechanism. During~an offline payment, however, common ledger consistency cannot be established at the time of the transaction. Offline CBDC designs must therefore aim for an appropriate balance between local integrity protection, subsequent synchronization, and, where necessary, the~detection and handling of protocol-conflicting or fraudulent~transactions.

In practice, the~OPF could be realized in the form of Smartphone applications, NFC Smartcards or dedicated devices with some storage and computation power, which can be used to store funds and to transfer funds to each other in close proximity, but~without reliance on e.g.,\ internet connectivity.
The attachment to the traditional banking system or an online CBDC is understood in a way, that these devices act as ``offline pockets'' and can be funded and de-funded when online (i.e., transfer of funds from bank account) or e.g.,\ connected to an ATM machine.
Thus, the~OPF of the CBDC is considered as a bearer-instrument implementation, whenever it follows the above funding mechanism~\cite{ecbfaqDE}. {(Note again that this is independent of the underlying technical realization of fund representation and settlement.)}

\begin{Definition}\label{def:transferabilityoffline}
    An OPF fulfills $\textbf{transferability}$ if it allows for consecutive transfers while offline. This means, if~user $\textbf{u}_1$ transfers a fund $\textbf{f}$ to user $\textbf{u}_2$, and~$\textbf{u}_2$ consequently transfers $\textbf{f}$ to $\textbf{u}_3$, all while being offline, user $\textbf{u}_3$ should be able to 
    consider the receipt of $\textbf{f}$ 
    as final to the same degree of certainty 
    as $\textbf{u}_2$ considered the 
    previous receipt of $\textbf{f}$ as final.
    Moreover, $\textbf{u}_3$ should be able to unconditionally prove ownership of $\textbf{f}$.  
\end{Definition}

In particular, it is not required technically that any user has to communicate with the TA at any stage of this chain of transfers.
From now on, we restrict our investigation to CBDCs which offer a transferable OPF, and~call such a system an \textit{offline CBDC}.
In the jargon of~\cite{Polaris}, we exclude ``Staged offline''-systems and focus
on ``Fully offline''-systems. {(We remark that there is the additional
definition of an 
``Intermittently offline''-system, which can be considered as a safety feature added to a ``Fully offline''-system which restricts the transferability e.g.,~to chains of a certain maximal length. As~this does not alter the CBDC design on a fundamental technical basis, we ignore this distinction further on.)}

We allow in principle for offline operation for an infinite time period. 
However, it is established in broad generality that with each consecutive payment conducted offline, the~data that needs to be transferred between counterparts inevitably grows in size~\cite{chaum1992}.
Moreover, it is clear that performing a chain of offline transactions without synchronizing with the TA in between amounts to sacrificing the security provided by the TA.
Hence, both for practical and security considerations, it is appropriate to assume that any user will try to ``go online'' from time to time. 
However, we will from now on focus on the functioning of an offline CBDC during offline periods.
Note that a given transaction protocol can, in~principle, differ depending on the connectivity status and we ignore the question of how an online transaction is handled (i.e., a transaction where at least one participant is able to interact with the TA).

\subsection{Threat Classification for Offline~CBDCs}

Let us assume an offline CBDC of the following type and in the following situation:
Any user is the legitimate bearer of their funds and behaved according to the CBDC's ruleset in the past. {(As a consequence of this assumption, threats against ``unforgeability'' of funds, 
i.e., against their authenticity guarantee by the TA,
are out of scope for this research.)}
We assume that all logic and storage connected with the CBDC usage (e.g., the~core functionalities for payments as well as privacy and security defining routines of the wallet software), are confined within some (computational) environment $\textbf{e}$.
Such an environment $\textbf{e}$ is often realized as a trusted execution environment (TEE). 
In practice, such an environment might be a dedicated smartcard chip, or~an embedded secure element in a smartphone, or, in its simplest form, areas of a computer's persistent memory, RAM and CPU where the wallet app is supposed to be stored, loaded and~executed.

In such a situation, we can distinguish two kinds of threats:

\begin{Definition}\label{def:SecurityThreatsFirstkind}
    A security threat of the first kind (relative to $(\textbf{u}, \textbf{f})$) is any threat emanating from a third party without access to $\textbf{u}$'s device or TEE $\textbf{e}$, directed at disposing over the fund $\textbf{f}$ belonging to $\textbf{u}$ without the consent of $\textbf{u}$.
\end{Definition}
    A security threat of the first kind is a threat to the \textit{Access Control Security} of the CBDC, i.e.,~it aims to compromise the security of a user's funds by trying to achieve unauthorized disposal of the funds without direct access to the device or the TEE.
    Thus, \textit{Access Control Security} is a major objective regarding the design of any CBDC system in order to ensure trust and adoption by citizens into the system.
    An example for a violation of \textit{Access Control Security} would be a (remote) attack of one user against another user, exploiting poor CBDC design or security vulnerabilities, resulting in stealing that users~funds.

\begin{Definition}\label{def:SecurityThreatsSecondKind}
    A security threat of the second kind (relative to $(\textbf{u}, \textbf{f})$) is any threat emanating from $\textbf{u}$, directed at fraudulently collecting exchange value from other CBDC users for $\textbf{f}$ without actually legitimately and permanently transferring $\textbf{f}$ to the counterparty.
\end{Definition}
\textit{Security against Depositor Misbehavior} is defined as the aim of safeguarding against threats emanating from fraudulent misbehavior of depositors themselves, i.e.,~users misusing the access to their own device or wallet. 
This objective is essential to guarantee the financial stability of a currency union adopting a CBDC, since it aims to protect against counterfeit money and, thereby, ensures trust in the value of a~currency.

It is important to remark that \Cref{def:SecurityThreatsFirstkind,def:SecurityThreatsSecondKind} are not exhaustive, i.e.,~there might be many more security threats. 
For example, we could consider the case where an attacker can gain (physical or remote) control over $\textbf{u}$'s device and exploit this control to steal funds. 
This would consist of an edge case between the first and the second~kind.

Moreover, a~security threat of the second kind might be indirect, e.g.,~by colluding with receiving complices, who then perform the final fraud with bona fide CBDC users.
As we are confining ourselves to an offline situation, attacks that rely on collusion with the TA or flaws in the online payment flow (i.e., flaws on the side of the TA's database or online settlement) are out of scope of the current~discussion.

Further details are presented in the following~section.

\subsection{Double-Spending and Funds~Recovery}

Let us define a particular attack situation of the second kind:
\begin{Definition}
    A \textbf{double-spending attack} is a fraudulent attempt of the bearer of one asset to give two different 
    users the false impression that they are both being transferred said asset.
    We exclude attempts that 
    rely on manipulation of the receivers' device or communication channel. {(For example, we exclude such attacks which rely on manipulating the receivers' device to show a balance or incoming transaction on the screen which is not actually within the CBDC system. The reader can think about a fake ``lookalike'' Wallet app which
    only gives the impression of a transaction, but~does not perform any CBDC-related operations at all. Comparable to fraud based on photoshopped bank statements, this cannot effectively be prevented by bank procedures but only by raising awareness of the public.)
    }
    \begin{itemize}
        \item A \textbf{trivial double-spending attack} is one which is invalid at first glance and will be rejected immediately by both the payment counterparty and the settlement logic of the TA.
        \item A \textbf{double-spending race attack} is one that can trick the two recipients of the same fund into believing (at least for a certain time-span) that they legitimately received the funds. Ultimately, however, the~double-spent will be uncovered at a later point, usually if all concerned parties go online again, e.g.,~by verifications through the TA. 
        \item A \textbf{strong double-spending attack} will pass all checks and, thus, remains undetected, also upon reconnection with the TA.
    \end{itemize}
    \end{Definition}

    \par{\textbf{Examples of double-spending in the context of Bitcoin.}}
    While there is a considerable difference between the conceptual mechanics of double-spending in a blockchain consensus situation (like Bitcoin) and in an offline CBDC situation (where the attack is targeted against various detection mechanisms), we decided to devote a small paragraph to the well-studied former situation:
    
    In the Bitcoin-world, a~simple attempt to spend a previously unspent transaction output (UTXO, c.f.~the definition in~\cite{bitcoin}) with an invalid signature or an attempt to spend a UTXO twice will result in a rejected transaction, thus will not result in a credit note on the recipient's side. 
    These kinds of \textit{trivial double-spending attacks} are unavoidable, pose not much harm and will therefore be ignored from now~on. 
    
    Also in the context of Bitcoin, a~user spending the same UTXO twice and propagating the conflicting transactions to different recipients or parts of the network conducts a 
    \textit{double-spending race attack}. The~attack succeeds if both recipients accept the payment before having checked for ultimate confirmation in the blockchain, and~an economical advantage arises for the attacker if both recipients hand over the counter-value (e.g., goods or services) to the attacker based solely on the existence of a signed transaction.   
    In collusion with a significant proportion of miners to withhold certain mined blocks for a certain time, this can be modified into a Finney attack. {(This theoretical scenario, named after Bitcoin pioneer Hal Finney, describes a malevolent actor, who pre-mines a block with an own transaction being spent to another address under his control, but~withholding the block in order to trick a counterpart into accepting the same funds on the basis of a different (legitimate) broadcasted but unconfirmed transaction. This attack can only harm Bitcoin participants which consider unconfirmed transactions as final and hand out countervalue in exchange before confirmative blocks have been mined and published.)} 
    
    However, a~successful \textit{strong double-spending attack} would imply practically insurmountable efforts (e.g., manipulating the majority of miners and nodes) which are equivalent to invalidating the whole Bitcoin-system. 
    Thus, for~practical purposes, Bitcoin is considered double-spending proof, provided that any recipient checks for a sufficient number of confirmations in the blockchain (e.g., six confirmations) before considering the receipt as \textit{final} and handing out the respective exchange value of the Bitcoin~transfer. 
    
    For the setting of this discussion, i.e.,~the behaviour of an offline CBDC during a period of offline usage, it is clear that the \textit{double-spending race attack} is the relevant threat condition, so we suppress the ``race'' from now~on.

There is a desirable feature of offline CBDCs which also seems worth mentioning:
\begin{Definition}
    An OPF allows for \textbf{funds recovery} if there exists a recovery mechanism for a user \textbf{u} to regain his funds \textbf{f} in case the environment \textbf{e} gets lost or destroyed.
\end{Definition}
    We remark that such a mechanism cannot rely on a proof, as~such a proof is not available in all cases (e.g., a user cannot prove whether its device vanished without a trace or whether he just hides his device). We furthermore remark that such a mechanism cannot rely on blocking a reportedly lost device, because~there is no way to enforce such a blocking on a constantly offline device used for offline~payments.

\begin{observation}
    An offline CBDC that is entirely immune against double-spending race attacks (despite the technical challenges of ensuring this immunity) cannot have a funds recovery mechanism.
    A similar observation is formulated in~\cite{bankofCanada2008bestbefore} as the following trilemma: We can have at most two out of three properties: “offline payment functionality,” “no double-spending” and “loss of funds not implied by device loss.”
    Note that even when an offline CBDC does not have the last property, it is already technically challenging to guarantee that no double-spending occurs, as~will become clearer in \Cref{sec:AlignObjectivesAndDesignElements}. 
\end{observation}

\section{Privacy and Data~Protection}
 \label{sec:Privacy}

If we consider, as~an example, the~situation in the European Union, the~relevant principles 
are detailed in 
the General Data Protection Regulation (GDPR), EU Regulation 2016/679
~\cite{EUreg}.
Although the GDPR does not explicitly define or mention the term “privacy,” it establishes a comprehensive framework for the protection of personal data, which is a core aspect of privacy in the European context. 
Article 4 defines ``personal data'', which means any information relating to an identified or identifiable natural person (i.e., data subject) and discussing its ``protection''. {(According to the Judgment of the Court issued by the European Court of Justice (ECJ) the 4th of September 2025, it has been  decided/stated  that pseudonymised data is not automatically personal data for all parties. Rather, its classification depends on whether the recipient can reasonably reidentify individuals, taking into account technical, organisational and legal factors, cf.}~\cite{ECJJudgement}{.)} 
Furthermore, Article 5  points out seven principles related to processing personal data, which closely build upon the \textit{Privacy by Design} philosophy: (i) \textit{Lawfulness, Fairness, and~Transparency}---processed lawfully, fairly and in a transparent manner in relation to the data subject; \mbox{(ii) \textit{Limitation on Purposes of Collection, Processing, and~Storage}---collected} for specified, explicit and legitimate purposes on not further processed in a manner that is incompatible with those purposes; furthermore processing for archiving purposes in the public interest, scientific or historical research purposes or statistical purposes shall not be considered to be incompatible with the initial purposes (iii) \textit{Data Minimization}---adequate, relevant and limited to what is necessary in relation to the purposes for which they are processed; \mbox{(iv) \textit{Accuracy}---accurate} and, where necessary, kept up to date; every reasonable step must be taken to ensure that personal data that are inaccurate, having regard to the purposes for which they are processed, are erased or rectified without delay; (v) \textit{Storage limitation}---kept in a form which permits identification of data subjects for no longer than is necessary for the purposes for which the personal data are processed; personal data may be stored for longer periods as the personal data will be processed solely for archiving purposes in the public interest, scientific or historical research purposes or statistical purposes; \mbox{(vi) \textit{Integrity and Confidentiality}---processed in a} manner that ensures appropriate security of the personal data, including protection against unauthorized or unlawful processing and against accidental loss, destruction or damage, using appropriate technical or organizational measures; (vii) \textit{Accountability}---there shall exist entities responsible for, and~able to demonstrate compliance with the principles above.
Importantly, the~GDPR also explicitly incorporates the concept of \textit{Privacy by Design}  and by Default in Article 25, requiring data controllers to implement appropriate technical and organizational measures to integrate data protection into processing activities from the outset. While \textit{Privacy by Design}  is a broader philosophy that advocates for embedding privacy considerations into the design of systems and processes, the~GDPR makes certain aspects of this approach a legal obligation for organizations handling personal data. In~summary, while GDPR compliance and \textit{Privacy by Design}  are distinct frameworks, the~GDPR both reflects and mandates key elements of \textit{Privacy by Design}, making them closely interconnected in~practice.

\textbf{Implication for CBDC Design.}
Starting from the above stated principles of the \textit{Privacy by Design} philosophy as well as the 
GDPR principles, specific privacy goals related to the realization of a CBDC are derived and summarized as:

\begin{itemize}
    \item  \textbf{User Anonymity towards the TA}: This goal concerns the ability of users to make payments without revealing their identity, that means the TA (for example: the central bank) does not know who is spending, where, or~when.
    To achieve \textit{user anonymity}, the~TA should validate and settle the transaction with minimized data collection, but~cannot link it to the user and is not able to derive personal information or payment patterns.
    This is in line with principle $2$ of \textit{Privacy by Design} or principle (iii) of the GDPR.
    In CBDCs, user anonymity, as~covered in principle $5$ of \textit{Privacy by Design}, refers to the protection of a user's identity throughout the lifecycle of a transaction—from withdrawal to payment—such that no party, including the TA, can associate a specific payment with a specific individual. 
    This objective is critical for ensuring that users can transact privately, mirroring the anonymity of physical cash.
    To achieve this, the~CBDC must avoid embedding any personally identifiable information in the digital currency or payment~protocol.

    \item \textbf{Transaction confidentiality}: This goal ensures that transaction-related information is protected, and~accessible only to authorized persons, preventing unauthorized access or disclosure of sensitive data to third parties.  
    To ensure strong privacy in a CBDC, both payment and balance \textit{confidentiality} must be preserved. 
    Note that this privacy requirement makes the validation of a transaction as well as possible additional features (like e.g.,\ fraud prevention) a highly non-trivial challenge, which will be addressed below and, in~more detail, in~\Cref{sec:AlignObjectivesAndDesignElements}.
    A user's wallet balance should be visible only to the owner, and~payment values should be known exclusively to the sender and recipient. 
    Identities of both parties must remain anonymous from third parties, including the TA. 
    When users make payments while offline, these transactions are temporarily stored on their devices and later uploaded once the user synchronizes with the TA.
    Even during this synchronization, privacy/confidentiality must be maintained, ensuring no sensitive data is exposed without consent, see principle $3$ of \textit{Privacy by Design}.

    \item \textbf{Non-Traceability}: As traceability refers to the ability to trace transactions back to users, \textit{non-traceability} in offline CBDCs refers to the inability to trace a specific transaction back to the identity of the user who performed it. This goal aims to ensure that
    no entity—including the TA—can retroactively determine who made a given payment. It is a foundational privacy feature designed to replicate the anonymity of physical cash, where users can spend money without creating a traceable digital footprint. In~offline settings, where transactions are conducted directly between users or merchants without a connection to the TA, achieving non-traceability requires mechanisms that prevent the embedding or leakage of identifiable information, as~an essential requirement according to principles $3$ and $5$ of the \textit{Privacy by Design} concept.

   \item \textbf{Unlinkability}: This principle is characterized by the ability to prevent transactions from being linked to each other or to the user. 
   In the context of CBDCs, \textit{unlinkability} refers to the inability to associate multiple transactions with the same user or to link different transactions to each other. 
   This goal is essential for preserving user privacy by ensuring that, even if a TA or any third party observes multiple payments, they cannot determine whether those transactions were made by the same person or received by the same person. 
   In an offline setting, where transactions occur without real-time connectivity to central infrastructure, unlinkability protects users from profiling, behavioral tracking, and~surveillance. 
   Unlinkability is fundamental to mimic the privacy-preserving properties of physical cash, where each transaction stands independently, offering no clues about the payer’s identity or payment history. 
   Moreover, it can be directly traced to principle $5$ of the \textit{Privacy by Design} concept.

\end{itemize}

We 
 remark that certain anonymity goals, like recognition of previously owned funds~\cite{chaum1992, Canard}, cannot be guaranteed on a theoretical level in the offline setting. (For example, Ref.~\cite{Canard} assumes an adversary with infinite computing power.)

\textbf{Main methods to privacy}: Several methods exist in the literature regarding how to achieve data protection, such as, for~example, pseudonymization, encryption or access controls. 
Here, we present two cryptographic methods that we consider the most suitable to meet the objectives described above, as~we will further elaborate upon below.
These methods are particularly robust and essential for the OPF of the CBDC, as~we discuss in detail in \Cref{sec:AlignObjectivesAndDesignElements}.
More specifically, the~first method is called \textit{Blind Signature Scheme} \mbox{(BSS) \cite{chaum1983blind}} which can be described as follows:
The content of the message is not visible by the signer, who still (as in regular cryptographic signatures) performs a signing operation on the blinded version 
of the message. While the signature of the unblinded message is verifyable, the~signer cannot determine that he previously signed that particular message.
The use of BSS can help providing privacy and anonymity, see also~\cite{goldwasser1996lecture} for more information.
The second method is called \textit{Zero-knowledge Proof} (ZKP) \cite{goldwasser1989zkp}. It is a cryptographic method by which one party can prove to another that a statement is true without revealing any information beyond the validity of the statement itself. 
In the context of CBDCs, BSSs and ZKPs would allow users to verify to each other that they are in possession of funds without disclosing their identity, balances or other privacy-relevant information.
For example, one can prove to the TA that a transaction is valid without revealing sensitive details, see also~\cite{beer2024payoffregulatedcentralbank} as an~example.

\begin{observation} In a CBDC system, 
compliance with the GDPR, in~line with the \textit{Privacy by Design}  philosophy, will be achieved if the 7 principles are formally defined and technically implemented within the system’s infrastructure.
Specifically, this means that the properties of anonymity, confidentiality, non-traceability, and~unlinkability are guaranteed not only through policies, organizational rules or law, but~most importantly through integrated technical mechanisms (e.g., cryptographic protocols, anonymization techniques, access restrictions, such as ZKPs, BSSs).
These methods inherently intend to prevent misuse both by the central TA and by users themselves.
Thus, these measures can ensure robust and verifiable privacy protection, independent of subjective trust in the central operator, leading to a real-world application of the \textit{Privacy by Design}  concept to the context of (offline) CBDCs.
\label{obs:privacyComplianceThroughCryptography}
\end{observation}

While Observation \ref{obs:privacyComplianceThroughCryptography} 
holds for both CBDC in an online setting as well as in an offline setting, we now revisit the OPF of a CBDC, formulated and discussed in \Cref{sec:problemdefinition}.
Table~\ref{tab:privacy-objectives} provides an overview of the privacy-related subgoals in an offline CBDC and illustrates how BSSs and ZKPs may contribute to achieve~them.

\begin{table}[H]
\caption{\textit{Privacy
 by Design}  goals for CBDCs design and how the cryptographic methods BSSs and ZKPs can be used to achieve \textit{Privacy by Design}  goals within the generic CBDC system with OPF defined in \cref{sec:problemdefinition}. }
\label{tab:privacy-objectives}
  \begin{tabularx}{\textwidth}{lL}
\toprule
\textbf{Sub-Goal} & \textbf{Exemplatory Application of the Identified Methods} \\ \midrule 
Non-Traceability & 
When combined with appropriate protocol and design choices (regarding the handling of e.g.,~transaction data, identifiers and metadata), ZKPs can help limit the information available for tracing values back to users or previous spends. BSSs can be used to limit traceability at issuance, if~combined with appropriate design and implementation choices. 
\\ \midrule
Unlinkability & BSSs can help unlink values from users at issuance, and~ZKPs can be used to achieve that offline transfers cannot be linked to one another or to the user, preserving unlinkability across multiple transactions independent of participants being online or offline. It is important to note that these features are not provided automatically, but~depend on adjuvant choices in the protocol design, e.g.,~by avoiding the use of linkable identifiers and metadata in the redemption process.
\\ \midrule
Confidentiality & ZKPs allow proving validity of transactions without revealing amounts, serials of wallets or (if in usage) tokens, or~user identities. Thus, they keep transaction details confidential from the TA and others. \\ ine
User Anonymity  & BSSs can help ensuring the TA cannot associate values to users, while ZKPs can be used to protect user identity during offline transfers and redemption. These features equally depend on suitable design and implementation choices and are not automatically guaranteed by incorporating BSSs and ZKPs.
\\ \bottomrule
\end{tabularx}

\end{table}

\section{Aligning Objectives and Design~Elements}\label{sec:AlignObjectivesAndDesignElements}

The associations developed in the course of this section should be understood as a structured design-level mapping rather than as a completeness or uniqueness result. Our analysis proceeds from the threat classification introduced above: for each objective, we identify the affected IT security principle, derive the corresponding technical requirement, and~then discuss design elements that address this requirement. Our mapping is therefore not intended to exclude complementary mechanisms or integrated protocol designs. Rather, it clarifies which countermeasures are primarily suited to address which threat class, and~where substitutions between design elements are~limited.  

From the discussions in the two previous sections, we focus on the following three key~objectives:
\begin{itemize}
    \item [-] \textbf{Access Control Security} is the set of measures that aims to protect systems and data from users according to threats of the first kind;

 \item[-] \textbf{Security against Depositor Misbehavior} is the approach of applying controls that deal with the detection, prevention, and~management of risks originating from malevolent users according to threats of the second kind;

 \item [-] \textbf{Privacy by Design} is a methodology that focuses on embedding privacy protections and data security into the system and processes from the very beginning.
\end{itemize}

{We} want to point out that these three objectives aim to specifically address the 
security and privacy challenges of CBDCs with OPF. There are equally important functional or usability-centered 
dimensions, e.g.,~scalability, latency, availability, regulatory compliance, or~financial stability which would constitute promising 
topics for complementary future~research.

In the following subsections we systematically relate these key objectives to concrete design elements and countermeasures.
The finding is that each objective is most directly addressed by a specific class of design elements.
We summarize this alignment of design elements and objectives in \cref{table:summary_obj_designelement}. 
An in-depth reasoning for the relationship between objectives and design elements can be found in \cref{subsec:ZKPs_for_privacy,subsec_sechardwarforSDM,subsec:sign_forACS}.
Overall conclusions and key takeaways of our analysis can be found in \cref{subsec_keytakeaways}.

\begin{table}[H]
\caption{\label{table:summary_obj_designelement}
Objective-countermeasure associations for \textit{Access Control Security}, \textit{Security against Depositor Misbehavior}, and~\textit{Privacy by Design}.}

  \begin{tabularx}{\textwidth}{LLL}
\toprule
\textbf{Objective} & \textbf{Technical Requirement} & \textbf{Countermeasure} \\ \midrule

\textit{Access Control Security}: Decrease vulnerability from threats of the first kind
& Authenticity of Transactions
& Digital Signatures based on Asymmetric Cryptography \\ \midrule

\textit{Security against Depositor Misbehavior}: Decrease vulnerability from threats of the second kind
& Integrity of the user's individual execution environment and wallet logic; integrity and confidentiality of wallet memory
& Adoption of \textit{secure hardware} \\ \midrule

\textit{Privacy by Design} on the level of transactions
& Confidentiality of user data, wallet states and transaction details while preserving the flow of information and evidence necessary for the functioning of the CBDC
& Complementing transaction data and cryptographic signatures by ZKP \& BSS \\ \bottomrule

\end{tabularx}
\end{table}

\subsection{Cryptographic Signatures for Access Control~Security}
\label{subsec:sign_forACS}
 In the context of CBDCs, with regard to \textbf{Access Control Security}, we address the level to 
which a CBDC design provides a user security with respect to threats of the first kind, i.e.,~how difficult it will be for user $\textbf{a}$ to steal the funds of another user $\textbf{b}$ (without having access either to $\textbf{b}$'s device/wallet or to the TA's system).
 
In this setting, any threat of the first kind translates to an attempt to impersonate the holder of a fund and to convince other participants that a transaction regarding this fund is initiated or approved by the legitimate user. Therefore, a~threat of the first kind can be reformulated as a threat to the authenticity of a transaction, e.g.,~in form of a transaction statement, which may manifest in various forms of impersonation or interception attacks, such as attempts to mediate communications without authorization or to unlawfully assume another’s~identity.

While it cannot be strictly excluded that fundamentally new digital authentication methods will
be developed one day, the~current state-of-the-art (see~\cite{nist_digitalsignaturestandard} by the National Institute of Standards and Technology for current US standards on digital signatures) can concordantly be described as follows:
\begin{observation} 
    Cryptographic signatures {(ECDSA is the current state-of-the-art standard and, thus, the~preferable cryptographic method of choice, and in~principle, it is possible to construct authentication mechanisms with also BSSs and ZKPs, which we consider then a subclass of cryptographic signatures; these methods are also suitable to address privacy concerns, as~discussed in} \Cref{subsec:ZKPs_for_privacy}; furthermore, ECDSA, for example, is based on
    asymmetric cryptography with elliptic curve arithmetic) represent the widely adopted, state-of-the-art digital method for establishing message authenticity that is publicly verifiable, non-repudiable, and~independent of pre-shared secrets, auxiliary communication channels, or~physical presence.
\end{observation}

We explain the necessity of the stated attributes in our CBDC scenario. For~practical purposes, all CBDC functionality—particularly transactions—should operate over the same digital communication path
(or, more precisely, all functionality shares the same 
communication protocol in the OSI application layer). 
This ensures that transactions do not depend on side channels such as phone calls for verification or the exchange of handwritten~signatures.

Moreover, while it is generally a good idea to shield communication, e.g.,~by cryptographic communication protocols, a~CBDC scheme should not 
rely entirely on the confidentiality of the communication between its participants. 
For example, consider previous man-in-the-middle attacks on protocols like TLS. 
Therefore, we assume the possibility of eavesdropping.
Clearly, it should be possible---in particular in the offline setting---that two users can transfer funds between one another without the possibility of prior individual authentication or sharing of secrets.
Finally, the~non-repudiability is justified by the requirement that a sender should not be able to unilaterally invalidate or challenge a payment~afterwards.

While there exist several general authentication methods {(Handwritten signature, Biometrics, Password/PIN-based authentication, Message Authentication Code, etc.)}, under~the above restrictions only digital signatures---based on asymmetric key cryptography---provide a verifiable and computationally non-repudiable proof of authorship of a given statement.
The soundness of cryptographic signatures and their usability for the described authentication application
is verified both in cryptography security theory and in practical application, and~no alternative
technique with comparable properties appears in scientific literature 
or IT practice (cf.~e.g.,~\cite{SMRB25, lynch2000authenticity, Katz2010, nist_digitalsignaturestandard}).

\subsection{Secure Hardware for Security Against Depositor~Misbehavior}
\label{subsec_sechardwarforSDM}

With regard to \textit{Security against Depositor Misbehavior} we formulate the objective of protecting a user $\textbf{u}$'s funds $\textbf{f}$ against threats of the second kind. 
Therefore, \textit{Security against Depositor Misbehavior} (with regard to $(\textbf{u}, \textbf{f})$)---in contrast to \textit{Access Control Security}---mainly
protects all other CBDC participants against misbehavior on $\textbf{u}$'s~side.

The essence of second-kind threats is that $\textbf{u}$  attempts to 
freely form, sign and transmit transaction statements on $\textbf{u}$'s funds and, thereby, violates the respective rulebook of the CBDC. 
There is an important~distinction:
\begin{enumerate}

\item A first example of a \textit{second-kind} threat is $\textbf{u}$ attempting to double-spend. Since neither the CBDC design nor its implementation provides such an option (e.g., no “double-spend button” in the wallet app), this would require $\textbf{u}$ to gain access to the wallet’s memory or execution logic in order to create, sign, and~transmit transaction statements that violate the CBDC rule set— that is, transaction statements which are invalid according to the CBDC's rule set.
The fraud arises when $\textbf{u}$ succeeds in obtaining value in exchange for funds backed by such an invalid statement, whose invalidity cannot be detected by the counterparty at the time of exchange (in particular, by~the receiver in an offline setting).

\item A derived form of the above second-kind attack is the \textit{cloning attack}. In~this scenario, we assume that $\textbf{u}$ cannot freely read or manipulate the wallet’s memory or execution logic, but~is able to produce identical copies—whether in hardware or software—of the wallet at a given state. Each cloned instance can then be used to transfer the same fund $\textbf{f}$ to different users, who are unable to distinguish whether the transaction originates from a legitimate wallet or from a cloned one that behaves~indistinguishably.

\item A further variation of the cloning attack is the \textit{rollback} attack. In~this scenario, we assume that $\textbf{u}$ can revert the wallet—including its memory and execution logic—to a previous state. This allows $\textbf{u}$ to spend the fund $\textbf{f}$ legitimately, then restore the wallet to the state immediately before the transaction and spend the same fund $\textbf{f}$ again with another beneficiary/payee. 
Note that in the offline situation the rollback attack involving spending of the same fund twice is, in~principle, identical to the \textit{replay} attack, where the same funds are used to form multiple, identical transactions.
\end{enumerate}

This highlights the following: 
Threats of the second kind are particularly relevant for offline CBDCs, since any double-spending attempt will usually be detectable upon reconnection, depending on the protocol. 
Consequently, the~practical advantage of double-spending arises from exploiting race conditions during offline periods. 
This observation underscores the centrality of double-spending as the specific security concern for \mbox{offline~CBDCs.}

On a theoretical level, second-kind threats target the integrity of transfer statements with respect to the CBDC ruleset, which typically defines a transaction as valid only if it spends a fund that has not been previously used.
This integrity of transfer statements can be reformulated into requirements of TEE integrity as well as memory integrity and confidentiality on the
side of the system hosting $\textbf{u}$'s~wallet. (It is crucial that the sensible parts of the memory and/or wallet logic (depending on the protocol) can be shielded against being read out or manipulated with. For~example, depending on the CBDC design, it must be prevented that $\textbf{u}$
can extract its own private key to authenticate double-spend transactions outside of the legitimate wallet logic. As~another example, it must be prevented that $\textbf{u}$ reads out the whole wallet application with all memory states in order to conduct a cloning attack.)

In this sense, second-kind threats ultimately target the system integrity of the individual TEE responsible for accounting for available funds, as~well as for creating, authenticating, and~broadcasting transaction statements—typically, the~secure element holding the user’s wallet~software.

We emphasize that during online connectivity, the~TA fulfills the role of anchoring trust with respect to transaction statement integrity in typical CBDC designs. 
So, one fundamental design objective is to replace the unavailable TA in the offline setting by anchoring the trust in the integrity of user's systems. 
Hence, a~user's system has to fulfill the role of TA towards other participants for integrity, particularly w.r.t.\ possible integrity breaches by the same user maintaining the system. 
In other words, protecting against second-kind threats to address \textit{Security against Depositor Misbehavior} is equivalent to protecting the system integrity of the user's individual TEE.
This insight comprises the notorious difficulty in offline CBDC design.
Fortunately, system integrity is a well-studied area, see~\cite{mukhopadhyay2014hardware} for an introduction to hardware security. The~existing literature allows us to formulate the following insights, see, e.g.,~\cite{mishra2025modernhardwaresecurityreview, 2023IEEEA, Jin_Intro_to_HW_sec, HCS2020,parikh2025survey,Boulifa2025faultinjection}:
\begin{observation}
  \textit{Secure hardware} and TEEs, in~general, are currently the most established practical class of mechanisms to robustly assure the integrity of execution environments, program 
  logic and memory areas against attackers with system access. These principles provide hardware-enforced isolation and tamper-proof separation of sensitive code and data from unauthorized access.
  In comparison to purely software-based principles, \textit{secure hardware} components establish a hardware root of trust, establish integrity e.g.,~through attestation mechanisms or secure boot procedures, and~protect against physical attacks and tampering. This ensures that the execution state, memory areas and program logic remain shielded, unaltered and verifiable throughout operation, as~consistently demonstrated by academic literature on trusted computing and in industry~practice.\\
  However, \textit{secure hardware} opens up a new attack surface and several successful attacks are described in the literature (side-channel attacks, timing attacks etc.), cf. e.g.,~\cite{roland2012relay,roland2012applying,Le2008SideChannel, 8141882, heriveaux2021defeating, tches-2019-29258}.

\end{observation}

It is important to note that there is little overlap with the previous topic. 
Thus, 
it is clear that \textit{secure hardware} cannot enhance \textit{Access Control Security}, as~an {outsider} has no opportunity to interact with the execution environment. 
On the other hand, weak cryptography can be exploited by both the user himself and outsiders alike (i.e., a user can forget about his advantages and behave like an outsider attacker).
However, if~we consider the common attack surface shared by the user and someone else versus the additional attack surface available to the user, it is evident that cryptographic methods primarily address the former.
In this context, the~role of \textit{secure hardware} is to minimize the advantage of the user compared to anybody else, in~the extreme to close this gap and to force a user trying to double-spend to tackle the cryptography like an outsider trying to double-spend someone else's funds, which should be designed as very difficult by~protocol.

\subsection{Zero-Knowledge-Proofs for Fundamental Privacy by~Design}\label{subsec:ZKPs_for_privacy}

Our next objective is to incorporate \textit{Privacy by Design} into a CBDC scheme. 
As discussed in \Cref{sec:Privacy}, this entails to ensure \textit{data protection} of user states and transaction statements against other participants, in~particular against the TA, the~payment counterpart and other eavesdropping/malicious~participants. 

At the same time, certain information must necessarily be transmitted to enable essential functions: for example, the~TA requires data for verification and settlement, while the payment counterpart needs information for verification and internal~accounting.

It is worth noting that, privacy cannot be treated as an isolated add-on. Instead, it requires redesigning the information exchanged (e.g., transaction statements) and its associated data (e.g., cryptographic signatures) so that they appear in a suitably data-minimized form. 
Such representations should disclose only what is strictly necessary for the functioning of the CBDC, while excluding personal or otherwise sensitive data. The~key points from the previous chapter and relevant literature are summarized as follows (cf.\ \mbox{e.g.,\ \cite{chaum1983blind,goldwasser1996lecture,goldwasser1989zkp,Garimella2024}).}

\begin{observation}
ZKPs and BSSs are ideally suited to provide confidentiality through minimization of required data within the transaction process, the~most fundamental process of the CBDC.
Thus, these two cryptographic techniques inherently allow to minimize the collection of privacy relevant data proactively in the transaction statements themselves.
When combined, these mechanisms, thus, support selective disclosure and unlinkability, offering mathematically rigorous, system agnostic mechanisms for achieving scalable, robust and preventive privacy in CBDCs.
Given the appropriate design of the CBDCs transfer statements, ZKPs and BSSs can be used to cover the first five principles of \textit{Privacy by Design} \cite{cavoukian2009privacy}, see the dedicated summary in \Cref{sec:Privacy} for~details.
 
\end{observation}
Note that other methods certainly can also be helpful to facilitate \textit{Privacy by Design} on other levels of a CBDCs.
The choice of (cryptographic) method, however, can strongly depend on the specific implementation of the CBDC.
For example, an~implementation based on a public blockchain could make use of stealth address protocols~\cite{liu2019key,buterin2025} to achieve unlinkability in the sense of \textit{Privacy by Design}.
This is possible through improving privacy issues associated to the attack surface of aligning users with public keys.
ZKPs and BSSs, however, minimize revealed data also between sender and receiver and, thereby, proactively and preventively minimize the attack~surface.

\subsection{Summary and Key~Takeaways}
\label{subsec_keytakeaways}
We can subsume and consolidate the information from the above subsections as follows, providing a unified overview that illustrates the relationships between design objectives, IT security principles, and~their corresponding design elements, with~respect to varying levels of~OPF, see \Cref{tab:summary_table}.

\begin{table}[H]
\caption{Summary 
 of design objectives, their association with design elements as countermeasures as well as alignment with affected IT-Security~Principles. \label{tab:summary_table} }
 
\begin{adjustwidth}{-\extralength}{0cm}
 \begin{tabularx}{\fulllength}{LLLL}
\toprule
\textbf{Objective} & Access 
 Control Security & Security Against Depositor Misbehavior & Privacy by Design\\ \midrule
Affected.
ITSec Principles & Authenticity & Integrity, \text{Authenticity} & Confidentiality \\ \midrule
Design Element & Cryptographic \text{Signatures} & Secure hardware & ZKP
\& BSS \\ \midrule
Offline \text{Specificity} \textsuperscript{1 
}  & None & High & Low \\ \bottomrule 

\end{tabularx}
\end{adjustwidth}
\noindent{\footnotesize{\textsuperscript{1} With “Offline Specificity” we refer to the following: The threat scenario described by \textit{Access Control Security} can occur in an online and an offline CBDC alike, and~the contribution made by the remedy “Cryptographic Signatures” does work equally well in the online and in the offline situation. In~contrast, 
the threat scenario of {a} depositor misbehaving---i.e.,~a threat of the second kind---differs very much, as~it is the offline scenario which opens
up the possibility for exploiting race conditions. When it comes to Privacy, the~placement is multifaceted: Generally speaking, ZKP techniques can be applied in a useful way {in order to prove} to a user receiving offline funds the authenticity of a previous transaction without revealing sensitive information. Also, note the analogy to physical cash payments where only payer and payee are involved in the transaction.}}

\end{table}

From the above discussion, the~following conclusions can thus be~drawn:
\begin{itemize}
    \item The three main objectives can be expressed in terms of the fundamental IT security principles: \textit{authenticity, integrity}, and~\textit{confidentiality}.

\item Each objective is most directly addressed by a specific class of design elements, and~this relationship is not arbitrarily interchangeable under the assumptions considered here. For~instance, in~an offline CBDC, the~threat of double-spending can be mitigated through the use of \textit{secure hardware}; however cryptographic signatures or ZKPs are not developed to and, thus, are not able to prevent double-spending in the offline situation. Similarly, an~offline CBDC design that does not incorporate ZKP and BSS techniques should be carefully investigated for privacy concerns, as~neither alternative cryptographic methods nor \textit{secure hardware} can substitute for ZKPs in providing privacy guarantees. This applies in particular to academic or commercial proposals which highlight e.g.,\ the use of cryptographic signatures. Such a proposal then deserves a thorough analysis to which objective which cryptographic method contributes, and~how other objectives -- which are not amenable to cryptographic  signatures -- are~addressed.

\item However, this approach should not be understood as a completely strict separation. In~particular, ZKP and BSS-based methods can incorporate the  role of cryptographic signatures and can be embedded on \textit{secure hardware}. So, ZKPs and BSSs are not (always) an independent add-on, which can be put on top of an offline CBDC to ensure privacy, but~rather made into an integral part of the protocol. However, this should not come as a contradiction, as~especially BSSs/ZKPs and cryptographic signatures are not strictly complementary to each other, but~rather ZKPs can be understood as sophisticated incarnations of cryptographic~signatures.

 \item Since double-spending is a primary concern when extending an online CBDC with an OPF, mechanisms that protect local wallet integrity have to act as an additional anchor of trust while the TA is not available. \textit{Secure hardware} is currently the most prominent practical candidate for fulfilling this~role.  
 
 \item Further research on cloning-resistant and rollback-resistant \textit{secure hardware} elements is essential, as~these constitute the primary enabler for secure offline~CBDCs.
    
\item Finally, special attention must be given to the producer and distributor of the \textit{secure hardware} elements. At~the level of integrated circuit (IC) design and manufacturing (e.g., at~the foundry), it is particularly challenging to eliminate the possibility of cloning or rollback capabilities. 
In most cases, the~manufacturer of the \textit{secure hardware} could, in~principle, execute double-spending race attacks on an arbitrary and potentially devastating scale.
\end{itemize}

\section{Examination of Existing Offline CBDC~Proposals}  \label{sec:proposal_examination}
Below, we analyze several previously published academic CBDC proposals with regard to the identified objectives of \textit{Access Control Security}, \textit{Security against Depositor Misbehavior} and \textit{Privacy by Design}, that are identified as crucial for CBDCs with OPF in \Cref{sec:AlignObjectivesAndDesignElements}; see \Cref{tab:assessment_proposalsI}.
Note that some of the chosen academic proposals are designed with respect to specific objectives in mind and are not necessarily intended for the real-world application.
Thus, the~assessment of these proposals shall not be understood as direct criticism of the respective works.
Instead, the~examination serves as an illustrative application of the proposed framework to selected offline-CBDC designs. It demonstrates how the framework can be used to organize and compare the treatment of the three objectives, without~constituting a formal test or validation of the framework itself.  
In Appendix \ref{app:assessment_cbdc_proposals}, the~literature search and selection process is described in detail, and, moreover, we explain why we consider the selected proposals a suitable representation of literature of payment proposals relevant to the design of CBDCs with~OPF.

The authors want to point out that the assessment is qualitative and non-weighted. It is not intended as a scoring or ranking scheme, since the relative weighting of the three objectives depends on the institutional and policy context of an envisaged CBDC deployment. Instead, we apply the following decision rules: An objective is marked as addressed as a main topic if the proposal explicitly discusses the corresponding threat and provides dedicated mitigation mechanisms. It is marked as tangentially addressed if relevant mechanisms are mentioned, but~the objective is not central to the proposal, i.e., only partially covered, or~treated as a black-box precondition. It is marked as not addressed if the available description does not mention suitable mechanisms for the respective objective at~all. 

We provide an overview over the individual assessments in \Cref{tab:assessment_proposalsI}.
A more detailed summary of the proposals as well as an elaborate assessment within our framework described in \Cref{sec:AlignObjectivesAndDesignElements} can be found in Appendix \ref{app:assessment_cbdc_proposals}. 
In each of the tables contained in Appendix \ref{app:assessment_cbdc_proposals}, there exist three columns dedicated to one of the three formulated objectives in our framework, in~addition to the summary above them. 
These columns summarize our examination of the respective work with respect to the associated objective and undermine the assessment result provided in \Cref{tab:assessment_proposalsI}.

Additionally, note the following couple of general observations:

\begin{table}[H]

 \caption{\label{tab:assessment_proposalsI}{{Assessment} 
 of selected references against the three~objectives. 
   \emph{Legend:} \Main{} = addressed as a \emph{main} topic;
  \Tang{} = \emph{tangentially} addressed;
  \None{} = \emph{not} addressed.   {Author names and their proposed schemes are highlighted via bold font.}} }
  
\small
\begin{adjustwidth}{-\extralength}{0cm}
    
  \begin{tabularx}{\fulllength}{cCCCC}
    \toprule
 \textbf{{Ref.}} & \textbf{Brief Description} & \textbf{Access Control Security} & \textbf{Security Against Depositor Misbehavior} & \textbf{Privacy by Design }\\ 
    \midrule 
~\cite{Chaum1990UntraceableEC} & \textbf{{Chaum et. al.} 
} describe \textbf{{Untraceable E-Cash}}, establishing a cryptographic system for E-cash with focus on privacy and non-tracability.  
    &  \Tang{}
    \begin{itemize}
        \item Access control enforcement through pre-image knowledge restriction, relying on the computational hardness of the RSA problem.
            \end{itemize}
        
         &  \Main{}
\begin{itemize}
        \item Double-spending prevention relies primarily on bank-side detection rather than hardware-based security.
        \item Coin blacklisting mechanism based on cryptographic redundancy revealed during double-spending.
        
        \end{itemize}
        
         &   \Tang{}
 \begin{itemize}
        \item Non-traceability guarantee for honest senders. 
         \item Privacy enforcement mechanism via blinded signatures combined with a cut-and-choose protocol during issuance and offline payment.
                 \end{itemize}
         \\

        \midrule

~\cite{Canard} & \textbf{{Canard and Gouget}} propose a \textbf{{Transferable E-Cash}} with focus on privacy and anonymity. Security objectives are formulated and assessed. Building upon ideas of~\cite{Chaum1990UntraceableEC}. 
    &  \Tang{}
 \begin{itemize}
   
        \item Ownership verification through proofs of knowledge of CL-signature and El Gamal-based security tag.
        \item Use of  El Gamal and CL-signature (strong RSA) assumptions, with~ZKP integration in the Santis–Yung framework.
                    \end{itemize}
                     & 
     \Tang{}
 \begin{itemize}
   
        \item Forgery prevention by digital signatures ensuring coin creation by trusted entities. 
        \item Double-spending detection via encrypted security tags, ZKP-based validity proofs, and~bank-side identification.
                \end{itemize}
                  & 
                    \Main{}
  \begin{itemize}
        \item Focus on anonymity and traceability aspects. 
         \item Two E-Cash schemes ensuring Full Anonymity (FA) \& Perfect Anonymity ($\text{PA}_2^\ast$) 
        \item Use of ZKPs and BSSs to achieve strong privacy
          \end{itemize}
          \\
        \midrule
~\cite{christodorescu2020twotierhierarchicalinfrastructureoffline} & \textbf{{Christodorescu et. al.}} describe a \textbf{{two-tier hierarchical trust infrastructure}} for CBDC with OPF. The~focus lies on instant completion with high throughput based on digital signatures.  &
   \Tang{}
 \begin{itemize}
        \item Addressed as "wallet security" by using asymmetric cryptographic signature inside a TEE
             \end{itemize}     & 
       
           \Main{}
                \begin{itemize}
        \item Double-Spending is mitigated by relying on a TEE which protects balance states and transaction signing from depositor misbehavior.
          \end{itemize}
        &
             \None{}
                \begin{itemize}
 \item Not discussed
                \end{itemize}
        \\ 
         \midrule
~\cite{Carvalho} & \textbf{{Videira}} describes a CBDC scheme with offline capabilities, where coins carry e.g.,~local blockchains for transaction tracing and expiration dates. & 
     \Tang{}
 \begin{itemize}
        \item Asymmetric public-key encryption shall be used, no further details of the particular method are provided
                    \end{itemize}
                     & 
        
              \Main{}
                \begin{itemize}
        \item Monotonous counters and two-stage approval with private key stored in TEE 
        \item Spending deadline limits offline periods
                 \end{itemize}
        
          & 
          
    \Tang{}
  \begin{itemize}
        \item Local blockchain exposes history of coin \item Possibility of pseudonymization briefly mentioned
               \end{itemize}  \\
        \midrule
~\cite{beer2024payoffregulatedcentralbank} &  \textbf{{Beer~et~al.'s}} CBDC scheme \textbf{{``PayOff''}} focuses on offline capability and strong privacy. Based on an account-based protocol using \textit{{secure hardware}
} and ZKP techniques. & 

    \Tang{}
  \begin{itemize}
        \item Cryptographic authentication with secret key \item Chosen schemes are exchangeable depending on  security requirements
               \end{itemize}
                & 
        
             \Main{}
                \begin{itemize}
        \item Identifies TEE on \textit{{secure hardware} 
} as crucial \item TA can cryptographically recognize counterfeit money and track origin of fraud
                 \end{itemize}
         & 
              \Main{}
                \begin{itemize}
         
         \item ZKPs and dependence proofs for secure, private offline payments
        \item Deanonymization only in case of fraud
                 \end{itemize}
         \\ 

     \bottomrule  
    
    \end{tabularx}
\end{adjustwidth}

\end{table}

\begin{itemize}
    \item In most proposals, the~\textit{Access Control Security} depends on a (often not explicitly specified) signature scheme, which is used as a black box. In~other (fewer) proposals, \textit{Access Control Security} is more integrated with ZKP methods, so the question of \textit{Access Control Security} cannot be independently reduced to the security of a given signature method in these cases.
    This hints at the fact that when applied to analyze an existing CBDC proposal, our three pillar approach cannot always be used in a direct manner, and~we have to account for valid proposals which use integrated measures to solve both \textit{Access Control Security} and \textit{Security against Depositor Misbehavior}.

    \item The usage of ZKPs aims to improve on privacy or provide \textit{Access Control Security}. Although~also the authors of this work consider ZKPs appropriate to achieve these objectives (see also \Cref{sec:AlignObjectivesAndDesignElements}), one has to consider that ZKPs might require rather long computations, which can take up to several seconds depending on the specific computation, especially for the limited resources available to a TEE such as on a secure element. This could be considered impractical for real-world applications, both in offline and online usage scenarios of CBDCs, depending on the circumstances and protocol~limitations. 

    \item Measures against depositor misbehavior, which do not incorporate \textit{secure hardware}, at~most provide a post-fraud traceability of second-kind threats including double-spending. Software/cryptographic measures on TEEs in combination with \textit{secure hardware} can render certain second-kind threats technically more involved, but~not prevent~them. 

    \item Throughout the selected proposals it becomes clear that double-spending through byte-by-byte cloning of the wallet can not universally
    be prevented by software or cryptographic primitives alone, given the transferable offline assumption and the threat model considered here.
    Shielding of the wallet execution environment and state can effectively
    be addressed through \textit{secure hardware} or a functionally equivalent trusted execution mechanism. However, there is never a guarantee that \textit{secure hardware} cannot be cloned. Nevertheless, cloning attacks can be made technically very difficult and potentially detectable through appropriate \textit{secure hardware} design, see also~\cite{mukhopadhyay2014hardware, mishra2025modernhardwaresecurityreview, 2023IEEEA, parikh2025survey,Boulifa2025faultinjection}. This is in line with the conclusions drawn in \Cref{sec:AlignObjectivesAndDesignElements}.

\end{itemize}

\section{Conclusions and Closing~Remarks} \label{sec:conclusion}

This paper explored how CBDCs can be designed to support secure and private offline payments. We focused on three main objectives that any CBDC with OPF should aim to achieve: \textit{Access Control Security}, \textit{Security against Depositor Misbehavior}, and~\textit{Privacy by Design}. By~examining these objectives in detail, we showed how specific technical elements (such as \textit{secure hardware}, cryptographic protocols, and~privacy-preserving methods) can act as countermeasures. Importantly, we found that these goals can be aligned with distinct design choices that do not necessarily interfere with each other, making it easier to reason about trade-offs and priorities in system design.
This framework for CBDC design evaluation was put into practice on five academic proposals for digital payment systems, including proposals for CBDC systems with OPF.
This analysis serves as an illustrative demonstration of how the three main objectives and their corresponding design elements can be used to structure a comparative analysis of offline-CBDC systems. It does not, however, constitute a formal validation of the framework's completeness, reliability, or~practical~usefulness.  

A key insight from our discussion is that, under~the assumptions considered here, software and cryptographic mechanisms alone cannot generally provide the same protection against offline double-spending as a TEE providing wallet state protection. \textit{Secure hardware} can enforce integrity at the wallet level and help detect or deter attempts to manipulate balances or payments, making attacks more difficult and harder to hide. It does not, however, replace the TA or an appropriate consensus and settlement mechanism for establishing consistency of the common ledger.
At the same time, ensuring privacy should better not rely only on trust from institutes or organizations. 
We argued that cryptographic methods (e.g.,~ZKPs and BSSs) are important to protecting user data and ensuring compliance with \textit{Privacy by Design}  principles.

Besides that, several aspects remain open for further exploration. While we highlight ZKPs and BSSs as powerful privacy tools, they can introduce computational overhead and may slow down transaction processing in offline transactions, an~aspect which is out of the scope of this work.
Similarly,  the~role of \textit{secure hardware} components and their interaction with the rest of the system (examples of issues are hardware trust chains,  dependencies, and~potential side-channel attacks) needs a deeper examination. 
Since security against attacks of the second kind, such as double-spending, cannot be guaranteed, it is also necessary to further explore methods for ex-post fraud resolution and operational governance, such as fraud detection or hybrid control mechanisms, in~order to reduce potential economic losses and weaken incentives for attackers.
Finally, our work focused mainly on conceptual design; that means in practice, trade-offs will likely arise between privacy, security, performance, and~usability, and~these will need to be evaluated through real-world~testing.

It is worth noting that other relevant dimensions of offline CBDC systems, including scalability, latency, availability, regulatory compliance, financial stability, usability, and~overall performance, were not considered as fundamental objectives of the present analysis. This choice reflects the specific scope of this work, which focuses primarily on the security and privacy properties of offline CBDC solutions. Moreover, several of these dimensions are closely related to architectural, policy, regulatory, and~implementation choices that are still under discussion in the context of CBDC design. Nevertheless, they represent important complementary requirements for the practical deployment of offline CBDCs and will be considered in subsequent work to provide a more comprehensive assessment of offline CBDC~solutions.

Potential future investigation could therefore look at performance benchmarking and scalability, the~integration of privacy-preserving protocols in real (or simulated) transactions, in~order to understand how offline CBDCs could become technically feasible and therefore practically usable, by~remaining at the same time not only secure and privacy-preserving, but~also efficient.
Additionally, the~authors envisage using the framework of this article to develop a modular CBDC approach, where each of the three objectives is addressed by a dedicated module with the appropriate design choice (as identified in this work) and is therefore amenable to individual analysis and encapsulated assurance w.r.t.\ each~objective.

\vspace{6pt} 


\authorcontributions{Research, Analysis, Writing—original draft---{D.-A.G., A.T. and M.W.} 
 The~authors contributed equally in all parts of the article in terms of literature review, adopted methodology, analysis and design. 
}

\funding{This research received no external~funding.}

\dataavailability{{No new data were created or analyzed in this study. Data sharing is not applicable to this article. }}

\acknowledgments{The authors would like to thank Silvio Petriconi and Jan Bönsel for several helpful discussions. 
During the preparation of this manuscript/study, the~authors used GPT 5.4 by \textit{OpenAI} for assistance with writing. During~the review process, the authors used GPT 5.5 and 5.6 by \textit{OpenAI} for assistance. The~authors have reviewed and edited the output and take full responsibility for the content of this publication.
}

\conflictsofinterest{The views expressed in this paper are those of the authors and not necessarily of the Deutsche Bundesbank or the Eurosystem. The~authors declare no conflicts of~interest. }

\abbreviations{Abbreviations}{%
~The following abbreviations are used in this manuscript:\\

\noindent 
\begin{tabular}{@{}ll}
BSS & Blind signature scheme\\
CBDC & Central Bank Digital Currency\\
ECJ & European Court of Justice \\
IC & Integrated circuit \\
FA & Full anonymity\\
OPF & Offline payment functionality \\
TA & Trust anchor\\
TEE & Trusted execution environment \\
UTXO & Unspent transaction output \\
ZKP & Zero-Knowledge proof
\end{tabular}
}


\appendixtitles{yes} 
\appendixstart
\appendix

\section{Detailed Comparison of Existing CBDC~Proposals} 
\label{app:assessment_cbdc_proposals}
\unskip
\subsection{Methodology for Proposal~Selection} \label{app:research-methodology}

To identify relevant approaches to offline Central Bank Digital Currency (CBDC) payments, we conducted a systematic literature search and selection process, whose steps are shown in Table~\ref{tab:literature_selection} and subsequently described.  The~initial search was based on the keywords “Offline Payment”, “Central Bank Digital Currency”, and~“Digital Banking”. The~resulting publications were screened based on their relevance to the specific security and privacy aspects investigated in this study. In~particular, we considered works addressing at least one of the following aspects: (i) \textit{Access Control Security}; (ii) \textit{Security against Depositor Misbehavior}; (iii) \textit{Privacy by Design}.

\begin{table}[H]

\caption{\label{tab:literature_selection}Literature search and selection~criteria.}
\small
  \begin{tabularx}{\textwidth}{lL}
\toprule
\textbf{Stage} & \textbf{Criteria} \\
\midrule
Initial search &
Offline Payment; Central Bank Digital Currency; Digital Banking. \\

First screening &
Access Control Security; Security against Depositor Misbehavior;
Privacy by Design. \\

Expanded search &
Electronic Cash; Digital Cash; E-Cash; relevant references. \\

Scientific relevance &
Peer-reviewed IEEE, ACM, and~Springer venues; recurring citations;
novelty and relevance for recent works. \\

Final selection &
Overall relevance to the research objectives and investigated
security/privacy dimensions. \\
\bottomrule
\end{tabularx}

\end{table}

Based on this first screening, we further examined the terminology used in the identified publications and their cited references. This resulted in the inclusion of additional relevant search terms, including “Electronic Cash”, “Digital Cash”, and~“E-Cash”. The~references of relevant publications were also examined to identify further proposals and foundational works that may not have been retrieved using the initial~keywords.

Finally, the~identified proposals were assessed according to their relevance to the scope and objectives of this study, with~particular emphasis on their treatment of offline payments and the aforementioned security and privacy aspects. As~an additional criterion, their scientific relevance and recognition were considered. This included the publication of the work in well-established peer-reviewed venues, such as IEEE, ACM, and~Springer journals and conferences, as~well as its recurrence in the references of other scientific studies. For~more recent works, for~which citation counts may not yet provide a meaningful indication of scientific impact, particular consideration was given to their novelty and direct relevance to the research scope of this~study.

This process resulted in the selection of five proposals for detailed analysis. The~selected proposals therefore represent approaches that were identified as particularly relevant to the research questions and the security and privacy dimensions considered in this work, rather than constituting an exhaustive catalogue of all existing offline-CBDC~proposals.

The subsequent analysis of the five proposals is intended as an illustrative application of the proposed framework. It demonstrates how the framework can be used to classify and compare selected offline-CBDC proposals. However, this application does not constitute a formal validation of the framework's completeness, reliability, or~practical~usefulness.

\subsection{Analysis of Five CBDC Proposals Using the Derived~Framework}

\begin{table}[H]
\small
  \begin{tabularx}{\textwidth}{LLL}
\toprule
\multicolumn{3}{c}{\textbf{{Chaum~et~al.'s Untraceable E-Cash} 
} \textbf{(1990)}
  \cite{Chaum1990UntraceableEC}}\\
 \midrule

\multicolumn{3}{p{\dimexpr\textwidth-2\tabcolsep\relax}}{This article proposes an early cryptographic protocol for digital payments \textsuperscript{{{1}}
}, via so-called ``electronic cash'', with a focus on anonymity and non-traceability. The~authors base their proposal on the RSA signature scheme to cryptographically issue digital coins to users. This general concept is augmented with a cut-and-choose methodology, yielding a protocol where coins are untraceable, but~double-spenders can be de-anonymized with high probability.
 The authors conclude their article with a protocol extension to allow for digital checks (an instrument resembling guaranteed checks like EuroChecks), where similarly the bank cannot trace the spending of a check, and~with a proposed extension that allows for the blacklisting of all coins of an exposed double-spender. We remark that the cut-and-choose approach limits the proposal to single offline spending of issued coins (i.e., the proposal is not transferable in the sense of Definition 
 \ref{def:transferabilityoffline}).
Note that this early proposal already fulfills some properties desired for modern CBDC solutions, like the identification of double-spending and non-traceability of honest users, whereas other features like transferability in the offline setting were not considered.}
\\
\midrule
\textbf{Access Control Security} & \textbf{Security against Depositor Misbehavior} & \textbf{Privacy by Design} \\ \midrule
The spending of coins owned by another user is prevented by restricting the knowledge of a cryptographic pre-image (the third root modulo a suitable composite number) only to the legitimate withdrawer. Essentially, the~questions of \textit{{Access Control Security}
} can be reduced to the computational difficulty of the RSA problem. 
Besides this, the~authors do not further discuss questions related to \textit{{Access Control Security} 
}. 
& The possibility of double-spending is not directly prevented (e.g., by secure elements, which were in their infancy at that point of writing), but~rather indirectly addressed by the threat that double-spenders can be identified with high probability by the bank. 
Corresponding policy and law are not a direct topic of~\cite{Chaum1990UntraceableEC}.
However, a~cryptographic redundancy in coins owned by a single user allows for the blacklisting of these tokens through the bank. 
This redundancy can only be uncovered whenever there is double-spending of a token through the respective user. 
& The proposed scheme guarantees \textit{non-traceability} for honest senders (but not for receivers, which are generally referred to as \textit{shopkeepers} in~\cite{Chaum1990UntraceableEC}).
Thus, privacy standards align well with \textit{Privacy by design} guidelines as long as no double-spending occurs.
Technically, the~non-traceability is realized by using blinded signatures in conjunction with a cut-and-choose protocol both during issuance and during offline payment.
\\ \bottomrule
\end{tabularx}
\noindent{\footnotesize{\textsuperscript{1} Note that earlier dated works~\cite{chaum1983blind, chaum1985security, chaum1988privacy} are also by Chaum and collaborators. These works provide the foundation for the work~\cite{Chaum1990UntraceableEC} discussed here. Thus, the~pioneering works by Chaum and collaborators can be viewed as the first ideas for digital payment.}}

\end{table}

\begin{table}[H]

\small
  \begin{tabularx}{\textwidth}{LLL}
\toprule
 \multicolumn{3}{c}{\textbf{Canard and Gouget's Transferrable E-Cash} \textbf{(2008)} \cite{Canard}}\\
 \midrule
\multicolumn{3}{p{\dimexpr\textwidth-2\tabcolsep\relax}}{This article gives an in-depth-analysis of various anonymity properties of transferrable E-Cash regimes.
\begin{frame}

\[
\begin{tikzcd}[
  ampersand replacement=\&, 
  row sep=0.8em,
  column sep=0.8em,
]
 \& \& \&
\text{PA}_1^\ast \arrow[ld,symbol=\not\Rightarrow] \\
\text{Weak Anonymity} \arrow[r,symbol=\Leftarrow] \&
\text{Strong Anonymity} \arrow[r,symbol=\Leftarrow] \&
\text{Full Anonymity} \& \Leftarrow\&
\text{Perfect Anonymity} \arrow[ul,symbol=\quad\Rightarrow]
\arrow[dl,symbol=\quad\Rightarrow]\\
 \& \& \&
\text{PA}_2^\ast \arrow[lu,symbol=\not\Rightarrow] 
\end{tikzcd}
\]
\end{frame}
It is argued that \textit{Perfect Anonymity} in its broad sense is not achievable, but~that an E-Cash protocol fulfilling \textit{Full Anonymity}, 
\textit{$\text{PA}_1^\ast$} and \textit{$\text{PA}_2^\ast$}
 exists (even though an efficient implementation is not constructed in the scope of this paper).
The authors elaborated a generic E-Cash proposal fulfilling Full Anonymity as well as a more specific proposal satisfying $\text{PA}_2^\ast$, which is based on  previous articles of the authors on Transferrable E-Cash as well as the work on Compact E-Cash~\cite{CHL}.}
\\
\midrule
\textbf{Access Control Security} & \textbf{Security against Depositor Misbehavior} & \textbf{Privacy by Design} \\ \midrule
The issue of \textit{Access Control Security} does not constitute the main focus topic of this publication. However, the~proposed $\text{PA}_2^\ast$-protocol connects ownership of 
a coin with providing a proof of knowledge of a
Camenisch--Lysyanskaya signature (a ZKP) created by the bank during coin issuance and of knowledge of the El Gamal-style computation of a security tag $\textbf{T}$. 
Thus, \textit{Access Control Security} depends on the security of El Gamal cryptography, which relies on the Diffie-Hellman Problem and on the security of the Camenisch--Lysyanskaya signature (as discussed in {Section~3.5}
 of
~\cite{CHL}) which can be reduced to the strong RSA assumption.
Besides this, it seems crucial before any real-world application to thoroughly study the security implications of the ZKP technique employed, in~particular in the specific usage in the proposed scope of the Santis-Yung metaproof framework.
 
 & \textit{Security against Depositor Misbehavior} is addressed similarly through cryptographic mechanisms, which prevent forgeability, i.e.,~limiting the ability to create coins to trusted entities through digital signatures, and~provide the possibility to identify double-spenders:
 Each spending procedure assigns a security tag, computed by the sender, to~the coin's metadata. 
 For consecutive spends, the~security tags are encrypted. Validity of the coin data is proven to the receiver  through computation of a ZKP.
 Banks receiving such a coin are able to decrypt the individual security tags to identify the culprit in case that double-spending occurs (by means of the \texttt{Identify}-routine).
 Nevertheless, no hardware security mechanisms suitable to eliminate double-spending or other second-kind threats with high confidence are provided or discussed.

&The article highly prioritizes anonymity and partial traceability aspects, building on the known concepts of Weak and Strong Anonymity.
Thus, the~article investigates and proposes two E-Cash schemes with the following properties:
\textit{1. Full Anonymity:} The proposed scheme fulfills FA if an  adversary with unlimited computing power and access to public keys of banks and users is not able to recognize a coin that he has already observed during a spending between two honest users (although he recognizes a coin previously owned by himself). 
\textit{2. Perfect anonymity:} In addition to FA, this adversary is not able to decide whether or not he has already owned a coin he is receiving. The~proposed scheme fulfills the variant $PA^\star_2$, where the adversary is slightly weaker.
Both proposals make use of ZKPs and BSSs to fulfill the rigorous privacy properties. 

\\ \bottomrule
\end{tabularx}
\end{table}

\newpage 

\begin{table}[H]
\small
  \begin{tabularx}{\textwidth}{LLL}
\toprule
 \multicolumn{3}{c}{\textbf{{Christodorescu~et~al.'s OPS protocol}  \textbf{(2020)}} \cite{christodorescu2020twotierhierarchicalinfrastructureoffline}}\\
 \midrule
\multicolumn{3}{p{\dimexpr\textwidth-2\tabcolsep\relax}}{In this proposal by Christodorescu et. al. from VISA research, the~authors describe a balance-based CBDC scheme making use of a two-tier hierarchical trust infrastructure with offline capabilities. The~proposed offline payment system protocol is designed to offer instant completion with high throughput, and~relies on cryptographic signatures created within a trusted application being executed inside a TEE in order to secure users' funds and protect against double-spending.}\\
\midrule 



 \textbf{Access Control Security} & \textbf{Security against Depositor Misbehavior} & \textbf{Privacy by Design} \\ \midrule
The \textit{Access Control Security} is embedded  within the broader concept of
 ``wallet security'': It is addressed by using asymmetric cryptographic signatures within the TEE, which in turn is used to authenticate and authorize wallet operations, e.g.,~payments, thereby ensuring that only legitimate users can execute sensitive transactions.
 It is worth noting that, in~the description of the proposed protocol, the~cryptographic method itself is not specified by 
 the authors. This allows for flexibility, e.g.,~to shift to another method afterwards based on a changing security assessment, as~security standards evolve, or~as new vulnerabilities are discovered.

& 
In this CBDC proposal, protection against Depositor Misbehavior is based on a TEE, which can safely execute code within  protected \textit{secure hardware}, in~alignment with the conclusions of \Cref{sec:AlignObjectivesAndDesignElements}. Users' devices are able to interact with the TEE through an interface. The~TEE provides protection against replay and rollback attacks through monotonically increasing counters, which are incremented after every round of communication between parties. Double-spending is supposed to be prohibited by the TEE, which stores the sender account's balance.  These protection measures require careful monitoring of the production process of \textit{secure hardware} for offline-compatible devices\textsuperscript{{1}} as well as registration at the respective TA of the CBDC system. Moreover, dedicated hardware measures to prevent the byte-by-byte cloning of the TEE storage must be guaranteed.
& This article does not provide anything specific on privacy and data protection, neither in terms of management, nor in terms of available technologies, nor fundamental principles, except~that it should adhere to standards.\\ \bottomrule
\end{tabularx}\par
\noindent{\footnotesize{{\textsuperscript{1}} Note that, for~example, smartphones come with their own dedicated \textit{secure hardware}, which is not necessarily accessible for central banks, which are the natural TA for a CBDC.}}

\end{table}

\begin{table}[H]
\small
  \begin{tabularx}{\textwidth}{LLL}
\toprule
 \multicolumn{3}{c}{\textbf{ Videira's CBDC protocol proposal} \textbf{(2023)} \cite{Carvalho}}\\
 \midrule
\multicolumn{3}{p{\dimexpr\textwidth-2\tabcolsep\relax}}{In this article, a~CBDC scheme with offline capabilities is proposed, building on concepts from Bitcoin's blockchain and confirmation protocols~\cite{bitcoin} as well as the usage of TEEs from~\cite{christodorescu2020twotierhierarchicalinfrastructureoffline}. 
 Funds can be held in two types of offline coins (``cold'' and ``hot''), and~each offline coin carries its own local blockchain, where fractional payments are realized as forks in the coin's chain.
 Another distinctive feature is that the coins carry one (cold coins) or two (hot coins) expiration dates, allowing for wallet recovery mechanisms in the latter case, reminding of the discussion in~\cite{bankofCanada2008bestbefore}. The~author explains in detail how to handle partial payments and payment abortions (e.g., due to communication errors), a~topic often ignored in other proposals.}\\
 

\midrule 

\textbf{Access Control Security} & \textbf{Security against Depositor Misbehavior} & \textbf{Privacy by Design} \\ \midrule
The proposed CBDC design combines online and offline functionalities within a two-tier permissioned distributed ledger structure, where the central bank serves as root authority and intermediary financial institutions operate as certified nodes. 
For all transactions, the~proposal relies on an asymmetric public-key encryption method. 
The method is not narrowed down, except~for the requirement that it shall be ``unfeasible to derive a private key knowing the public key''. 
The author describes a public key infrastructure, clarifying the interplay with certificates of the banks and the central bank. 
Moreover, the~author suggests the usage of a double-signature authentication of transactions (one key stored on the TEE and one in the user's possession), as well as an additional authentication layer, as~PIN codes or biometric identifiers to~protect local wallets and prevent unauthorized access, establishing a multi-factor security model. 
& 
The prevention (mitigation) of double-spending is one of the main objectives of the proposal. 
In the first place, it relies on monotonous counters and a two-stage approval mechanism, where the wallet private key is stored inside a TEE, shielded from the user (similar to~\cite{christodorescu2020twotierhierarchicalinfrastructureoffline}).
This is complemented by incorporating a spending deadline, thus limiting the time intervals between online synchronizations and, thus, also the damage of double-spending. The~author discusses the threat of a brute-force attack against the TEE and a replay attack in detail. Additionally, the~author suggests establishing security thresholds (such as holding limits) to limit the possible damage, optionally based on individual user risk profiles. 

&
Privacy considerations are not central to the proposed design. 
The concept of a local blockchain for each coin inherently exposes transactional history to future holders and to the institutions involved in coin issuance and redemption (the latter is not detailed in the paper).
The author recognizes this limitation and proposes randomizing users’ public keys. 
While this approach may reduce direct linkability between transactions, it does not ensure full unlinkability or untraceability, leaving risks of profiling or deanonymization.
Moreover, the~framework explicitly values traceability as a design feature. 
Each coin’s blockchain is intended to support auditability and regulatory oversight, allowing us to track the history of coin ownership as a tool against money laundering and terrorism financing. 
In general, a~comprehensive assessment of \textit{Privacy by Design} principles is missing, which in turn would be essential to balance the issue between individual privacy and transparency.
\\
 \midrule
\multicolumn{3}{p{\dimexpr\textwidth-2\tabcolsep\relax}}{\textbf{Remark:} Note that the author suggests the concept of a ``dynamic chain'' to further enhance CBDC security.  
 The users' device continuously mines the local blockchain of a coin using a proof-of-work algorithm. 
 However, we think that this suggestion does not directly contribute to any of the security objectives. 
 Both from a theoretical perspective (i.e., how the dynamic blockchain could provide any benefit if vulnerabilities in the signature mechanism or TEE exist) and from a practical perspective (i.e., how the mining power of a cell phone could prevent an attack by a well-resourced threat actor), a~description and justification of the benefits of this approach would have been necessary. } \\
\bottomrule

\end{tabularx}
\end{table}

\begin{table}[H]
\small
  \begin{tabularx}{\textwidth}{LLL}
\toprule
 \multicolumn{3}{c}{\textbf{Beer~et~al.'s PayOff} \textbf{(2026)} \cite{beer2024payoffregulatedcentralbank}}\\
 \midrule
\multicolumn{3}{p{\dimexpr\textwidth-2\tabcolsep\relax}}{In this article, Beer~et~al.~propose a CBDC design which is highly adapted to the specifications of the Digital Euro project, focusing on comprehensive offline capability, strong privacy and enforcement of holding limits. The~authors describe an account-based protocol which makes crucial use of both \textit{secure hardware} and ZKP techniques. Towards the end, a~detailed discussion of performance figures and message sizes is given, together with pseudocode.}\\
 
\midrule

\textbf{Access Control Security} & \textbf{Security against Depositor Misbehavior} & \textbf{Privacy by Design} \\ \midrule
 In Beer et. al.'s proposal, \textit{Access Control Security} is explicitly defined as a security goal Theorem 7 (Payment Authorization). As~explained in {Section~5.2.1}
, 
 a fraudulent payment on behalf of another user requires knowledge of that user’s secret key or the blinding factor of the state commitment, both of which remain confidential to the user. Thus, such a breach of \textit{Access Control Security} can be reduced to a 
 vulnerability in the used ZKP system or commitment scheme.
ZKP techniques are used in a highly integrated way in the proposed payment process, as~they are used to confidentially confirm correct account updates in offline payments.
This necessitates a thorough security analysis of the used ZKP and commitment methods preceding a possible real-world application.
The authors use specific choices for the ZKP framework, signatures and hash functions (as outlined in {Section 6}) for their performance evaluation, but~the proposed protocol is not tied to these choices and could easily be adapted to changing security requirements.

 & 
 The article defines \textit{Security against Depositor Misbehavior} as Theorem 9 (\textit{Money and holding limit integrity}). It identifies \textit{secure hardware} elements enabling trusted code execution as a key component for offline CBDC systems. However, no secure element guarantees complete protection—hardware can still be compromised, enabling double-spending and counterfeit fraud. The~proposed solution addresses this by allowing \textit{de-anonymization} in double-spending cases.

Each offline transaction is accompanied by additional data (double-spending tags), which enables---if fraud occurs---a de-anonymization of the double-spending sender of the transaction. 
This allows the TA to compensate the victim in exchange for data necessary for the identification 
of the double-spender.
Consequently, the~proposal enhances the robustness of CBDCs relying on \textit{secure hardware}, though~it also increases message sizes during extended offline sessions and may raise additional privacy concerns.

& Only the participants of a payment should know its details, while TAs must not be able to track offline transactions or wallet states if users follow the rules. Payments from separate offline sessions shall stay unlinkable, cf.~Theorems 1-3. 
Still, the~system is able to reveal counterfeiters: if double-spent money is detected on reconnection, the~
victim shall reveal their identity to allow tracing the chain of transactions until the double-spender.
Thus, the~design balances \textit{privacy} and \textit{accountability} through a controlled \textit{de-anonymization} process (see Theorems 4-6 and Theorem 8).
The solution relies on \textit{ZKPs} for secure, private offline payments. 
Each user keeps a cryptographic state recording balance and transaction history without exposing details. 
Offline payments use ZKPs to prove balances and counters update correctly from valid prior states, preventing double-spending without revealing transaction data. 
Dependency proofs let users perform multiple offline payments before reconnecting by separating state correctness from signature verification. 
Upon reconnection, the~central bank validates all ZKPs and dependencies, signing legitimate states and flagging fraudulent ones.\\
\bottomrule

\end{tabularx}
\end{table}

\begin{adjustwidth}{-\extralength}{0cm}
\reftitle{References}

\PublishersNote{}
\end{adjustwidth}
\end{document}